\documentstyle[12pt,a4]{article}
\newcommand{\PR}{Phys.\ Rev.\ }
\newcommand{\PL}{Phys.\ Lett.\ }
\newcommand{\NP}{Nucl.\ Phys.\ }
\newcommand{\ZP}{Z.\ Phys.\ }

\def\S{${\tilde \sigma}[\c]$}
\def\s{\tilde{\sigma}[\c]}

\def\J{$J/\psi$}
\def\j{J/\psi}
\def\P{$\psi'$}

\def\U{$\Upsilon$}
\def\u{\Upsilon}
\def\c{c{\bar c}}

\def\lsim{\raise0.3ex\hbox{$<$\kern-0.75em\raise-1.1ex\hbox{$\sim$}}}
\def\gsim{\raise0.3ex\hbox{$>$\kern-0.75em\raise-1.1ex\hbox{$\sim$}}}

\newlength{\abstwidth}
\begin{document}

\sloppy

\pagestyle{empty}

\begin{flushright}
CERN--TH/95--75
\end{flushright}

\vspace{\fill}

\begin{center}
{\LARGE\bf Production of heavy quarks\\[1ex] and heavy quarkonia}\\[2ex]
{\Large Gerhard A. Schuler
} \\[2ex]
{\it Theory Division, CERN} \\[1mm]
{\it CH-1211 Geneva 23, Switzerland}\\[1mm]
{ E-mail: schulerg@cernvm.cern.ch}
\end{center}

\vspace{\fill}

\begin{center}
{\bf Abstract}\\[2ex]
\begin{minipage}{\abstwidth}
Uncertainties in the next-to-leading-order calculations of
heavy-quark ($Q$) production are investigated. Predictions for
total cross sections, single-inclusive distributions of heavy quarks
and heavy flavoured hadrons, as well as for $Q\bar{Q}$ correlations, are
compared with charm and bottom data.

$\quad$ The description of heavy-quarkonium production requires a separation
of the short-distance scale of $Q\bar{Q}$ production, which is set
by the heavy-quark mass from the longer-distance scales associated
with the bound-state formation. Various factorization approaches are
compared, in particular with respect to the different
constraints imposed on the colour and angular-momentum states of the
$Q\bar{Q}$ pair(s) within a specific quarkonium state. Theoretical
predictions are confronted with data on heavy-quarkonium production
at fixed-target experiments and also at $p\bar{p}$ colliders, where
fragmentation gives the leading-twist cross section in $1/p_T^2$ and
$1/m^2$ at high transverse momentum.
\end{minipage}
\end{center}

\vspace{\fill}
\noindent
\rule{6cm}{0.4mm}

\vspace{1mm} \noindent
${}^a$ Heisenberg Fellow.
\hfill\\[1ex]


\noindent
CERN--TH/95--75\\
March 1995

\clearpage
\pagestyle{plain}
\setcounter{page}{1}

\section{Production of heavy quarks}
\subsection{Theoretical status}
Total cross sections and single-inclusive distributions
of open heavy-quark production have been calculated in
perturbative QCD (pQCD) to next-to-leading-order (NLO) accuracy
for essentially all high-energy reactions,
hadroproduction \cite{NDE88},
photoproduction \cite{EN89},
leptoproduction (deep-inelastic lepton--nucleon scattering) \cite{LRSN93},
$\gamma \gamma$ collisions \cite{DKZZ93},
and deep-inelastic $e \gamma$ scattering \cite{LRSN94}.
In the case of photo- and hadroproduction, the NLO calculations
have even been implemented in a Monte Carlo programme \cite{FMNR93},
such that also double-differential distributions can be studied.
Attempts are ongoing to develop such an exclusive programme also
for deep-inelastic scattering \cite{Harris}.

In pQCD, the differential cross section of heavy-quark ($Q$) production
in the collision of hadrons $A$ and $B$ is given by
the factorized expression
\begin{eqnarray}
\lefteqn{
d\sigma[AB \rightarrow Q\bar{Q}X](p_A,p_B) =
}\nonumber\\ & & \!\!\!\!\!\!\!\!\!\!\!\!
 \sum_{i,j}\int dx_1\ dx_2 f_{i/A}(x_1,\mu_F^2)\ f_{j/B}(x_2,\mu_F^2)
   d\widehat{\sigma}[ij \rightarrow Q\bar{Q} X']
(x_1 p_A,x_2p_B,\mu_F,\mu_R)
\ .
\label{ABcross}
\end{eqnarray}
Here $i$, $j$ represent the interacting
partons (gluons, light quarks and antiquarks)
and the functions $f_{i/A}(x,\mu_F^2)$ are their number densities,
the parton distribution functions (PDF)
evaluated at momentum fraction $x$ and factorization
scale $\mu_F$. The short-distance cross section
$\widehat{\sigma}$ is calculable as a perturbation series in
$\alpha_s(\mu_R)$ where the strong coupling constant is evaluated at
the renormalization scale $\mu_R$. Leading-order (LO) diagrams
for heavy-quark hadroproduction are shown in Fig.~1.

The obvious question is, of course, whether pQCD does
describe the experimental data on charm and bottom production.
To answer this question the uncertainties in the theoretical
predictions have to be investigated.
These arise from four main sources:
\\[1ex]
1. The heavy-quark mass $m_Q$.
 Conservative ranges are
 $1.2 < m_c < 1.8\,$GeV and $4.5 < m_b < 5.0\,$GeV.
\\[1ex]
2. The size of unknown higher-order corrections. An estimate may
 be obtained by varying the factorization scale $\mu_F$ and the
 renormalization scale $\mu_R$ around their `natural' value
 $\mu_R = \mu_F = m_Q$, say between $m/2$ and $2m$.
 In the case of the charm quark, one has to note that
 most PDF $f_{i/H}(x,\mu_F^2)$ are not valid for $\mu_F\ \lsim\ 2\,$GeV
 and that the running of $\alpha_s(\mu_R^2)$ is no longer purely
 perturbative for $\mu_R\ \lsim\ m_c/2$.
 Proper error estimates are therefore not as easy as in the case
 of the $b$ quark (see below).
\\[1ex]
3. The value of the QCD scale $\Lambda$ and the shape of the PDF.
 Since these are strongly correlated
 (in particular $\Lambda$ and the gluon density), the choice of
 different $\Lambda$ values (in $60 < \Lambda_5 < 300\,$MeV, say)
 in the partonic cross sections
 should ideally be accompanied by parametrizations of PDF
 that have been fitted with those values of $\Lambda$.
 However, LEP data (and also the $\tau$ hadronic width)
 favour $\Lambda$ values that are almost a factor of two larger
 than is found in fits to deep-inelastic scattering data. In turn,
 one does not cover the full range of uncertainty in $\Lambda$
 when restricting to the available PDF parametrizations. In particular,
 the cross sections may be larger since larger values of $\Lambda$
 imply larger values of $\alpha_s$. In the absence of fits
 with $\Lambda$ frozen to the desired values one may choose \cite{FMNR94}
 to simply change the value of $\Lambda$ in the partonic cross sections.
\\[1ex]
4. Non-perturbative effects such as heavy flavoured hadron formation
 and the intrinsic $k_T$ of the incoming partons.

\subsection{Low-energy total cross sections}
In the case of charm production it turns out that the biggest
uncertainty arises from the variation of $m_c$.
In ref. \cite{FMNR94} it is found that
predictions of the lowest and largest values of
the total charm cross section in hadroproduction
may differ by up to a factor of $10$ from the central value
(i.e.\ the highest value is a factor of $100$ larger than the lowest one)!
Predictability substantially improves
when increasing the quark mass: for $b$ quarks, the
theoretical error is reduced to a factor of $\pm 2$--$3$.
Both for charm and for bottom, the predicted total cross sections
are in reasonable agreement with recent data (Fig.~2);
the charm data are claimed to be
consistent with a charm-quark mass of $1.5\,$GeV \cite{FMNR94}.

The uncertainty bands of the theoretical predictions in Fig.~2
represent the {\it maximum variation} of the cross sections in
a parameter range defined as follows \cite{FMNR94}. The scale
$\mu_R$ was varied in $[m_Q/2,2m_Q]$ and $\Lambda_4$ in
$[100,300]\,$MeV in order to account for a range of uncertainty
in $\Lambda$ wider than is provided by PDF fits.
However, only two particular sets of PDF parametrizations
(one for the proton and one for the pion \cite{HMRS})
were chosen with $\Lambda_4=190\,$MeV.
Hence the correlation between $\Lambda$ and the (nucleon and pion)
PDF was not properly taken into account. The mere variation of $\Lambda$
overestimates the associated uncertainty.

The scale  $\mu_F$ was also varied in $[m_b/2,2m_b]$
in the case of bottom, but
kept constant at $2m_c$ in the case of charm since the chosen PDF
parametrizations \cite{HMRS} are not valid at smaller scales.
The choice $\mu_F=2m_c$ maximizes the values of the PDF
for most values of $x$ and in turn of
the charm cross sections.
Moreover, PDF decrease (eventually non-perturbatively)
as $\mu_F^2 \rightarrow 0$.
Finally, the lowest allowed value of $\mu_R$ ($m_c^{min}/2 = 0.6\,$GeV),
together with the largest value of $\Lambda$ ($300\,$MeV),
gives an $\alpha_s$ value of $1.21$, which is certainly beyond
QCD perturbation theory.
In conclusion, the error bands in
ref.~\cite{FMNR94} are fairly pessimistic, in particular for charm,
and values of $m_c$ smaller than $1.5\,$GeV are favoured by the data.

The question whether the NLO calculation can accommodate the
hadroproduction data for $m_c=1.5\,$GeV was also addressed
in ref.~\cite{Mcgaughey}. The cross section for the choice
$\mu_R = \mu_F = m_c = 1.5\,$GeV
(and the PDF parametrization GRV \cite{GRV} that extends
down to $\mu_F \sim0.6\,$GeV) lies a factor of 2--3 below the data
(Fig.~3). A word of caveat on the data shown in Fig.~3 on the total
$c\bar{c}$ production cross section from $pp$ and $pA$ interactions
is in order.

First, in fixed-target experiments
mostly single charmed mesons in the region $x_F > 0$
are measured. Additional, poorly known  information is needed
to convert $\sigma(D/\bar{D})[x_F>0]$ into $\sigma[c\bar{c}]$.
So far only premature extractions of $\sigma[c\bar{c}]$ exist,
obtained by simple assumptions
on $\sigma(\Lambda_c)$, $\sigma(D_s)$, the $x_F$ distributions outside
the measured range, in particular in $pA$ collisions, the distortion
of the $x_F$ distributions due to non-perturbative effects, etc.
This explains (at least partly) the spread in the low-energy
($\sqrt{s}\ \lsim\ 40\,$GeV) data shown in Fig.~3.

Second, the ISR data ($53 < \sqrt{s} < 62\,$GeV),
inferred from lepton measurements in coincidence with a reconstructed
charmed hadron, are presumably too large due to the
assumed shape of the production cross sections:  flat
distributions in $x_F$ for the $\Lambda_c$ and $(1-x_F)^3$ for the $D$.
Restricting to the most recent experiments only consistency is found
\cite{FMNR94,Mcgaughey} between the data (Figs.~2 and~3)
and these data agree rather well with $\mu_R=\mu_F=m_c = 1.3\,$GeV
in the NLO calculation using the GRV PDF \cite{GRV}.
The result is similar (Fig.~3) with the MRS distributions \cite{D0}
and $\mu_R=\mu_F = 2m_c$ where now the large $\mu_R$ requires
an even smaller charm-quark mass, $m_c \sim 1.2\,$GeV.
Note that such small values of $m_c$ suggest that the bulk of the
total cross section comes from invariant masses smaller than $2m_D$.

The theoretical uncertainty is smaller in photoproduction:
there the Born cross section starts at $O(\alpha_s)$, compared
to $O(\alpha_s^2)$ in hadroproduction, and also only a single parton
density enters.
For example, the charm cross section can now be calculated
up to a factor of about $3$ \cite{FMNR94}. The theoretically
better-controlled $b$-quark cross section has unfortunately not
yet been measured in photoproduction; the forthcoming measurement
in $\gamma p$ collisions at HERA is eagerly awaited.

Additional tests of heavy-quark production
will soon also be possible in electroproduction
of heavy quarks at HERA and in electron--photon collisions at LEP2.
The most reliable predictions can be made for heavy-quark production
in $\gamma\gamma$ collisions. Even for $c$ quarks, the total uncertainty
is only about $\pm (30$--$35)$\% \cite{DKZZ93}.
The experimental situation is so far
controversial: taken at face value, the cross section seems to fall
when going from Tristan \cite{KEK} ($\sqrt{s}_{ee} \sim 60\,$GeV) to
LEP1 \cite{Aleph} ($\sqrt{s}_{ee} \sim 90\,$GeV)!

\subsection{Single-inclusive distributions}
Single-inclusive distributions can be calculated in pQCD by folding
the (NLO) cross sections of heavy-quark production with the
corresponding (NLO) fragmenting function, e.g.\ $c \rightarrow D+ X$.
In general, the $x_F$ and $p_T$ spectra of the heavy-flavour hadron
will be softer than those of the mother heavy-quark.
In the case of charm hadroproduction, the measured $D$-meson
$p_T$ distribution is considerably harder than the one
calculated in NLO perturbation theory.
Since the c.m.s.\ energies are small ($\sqrt{s} = 20$--$40\,$GeV),
it appears unlikely that the discrepancy can solely be resolved by
yet higher-order corrections. Rather the description of
differential charm distributions requires the inclusion of
non-perturbative effects. These may arise from the
intrinsic transverse momenta $k_T$ of the incident partons
and/or higher-twist corrections in the hadronization.

The effect of intrinsic $k_T$ is well established in the
Drell--Yan process and its inclusion gives also reasonable
agreement \cite{FMNR94} for $D$-meson $p_T$ distributions.
In fact, photoproduction
data are reproduced with $k_T$ values as they are also observed in the
Drell--Yan process, i.e.\ $\langle k_T^2 \rangle^{1/2} \sim 0.7\,$GeV.
On the other hand, hadroproduction data require values that are
as large ($\langle k_T^2 \rangle^{1/2} \sim 1.4\,$GeV)
as the charm-quark mass. Further studies are certainly needed,
in particular on the strong correlation of $k_T$
with $m_c$ (larger $m_c$ yields harder $p_T$ spectra and, in turn,
allows smaller $k_T$ values).

Distributions in $x_F$ have so far only be measured for charmed
hadrons and do not at all agree with the pQCD
predictions (they are too soft), even after
inclusion of the $k_T$ effects \cite{FMNR94}.
Indeed, an agreement must not be expected
in the first place, as factorization, underlying the perturbative approach,
is valid only at large $p_T$ and hence breaks down at large $x_F$.
Additional non-perturbative contributions associated with the
hadronization become important:
The enhanced production of $D$-mesons whose light valence quark is of the
same flavour as one of the valence quarks in the beam hadron
(leading-particle effect) and the dragging of charm quarks in the
colour field of the beam fragments (colour-drag effect).
Indeed, a fragmentation model that contains these effects,
such as the Lund string fragmentation one, successfully describes the
observed spectra. An alternative explanation \cite{Brodsky92}
invokes new production mechanisms for hard processes at large $x_F$ through
terms that are formally higher-twist effects $\propto 1/m_Q^2$ but
are enhanced by inverse powers of $1-x_F$.

\subsection{$Q\bar{Q}$ correlations}
In leading order, pQCD predicts that the heavy quark and heavy antiquark
are produced exactly back-to-back (Fig.~1),
implying $p_T(Q\bar{Q})=0$ and
$\Delta \phi = \pi$, where $p_T(Q\bar{Q})$ is the transverse momentum
of the pair and $\Delta \phi$ the angle between the projections of the
momenta of the pair onto the transverse (w.r.t.\ beam axis) plane.
NLO corrections, as well as non-perturbative effects, can cause a
broadening of these distributions. The question whether NLO predictions
can account for the available experimental data in hadro- and photoproduction
of charmed particles was addressed in ref.\ \cite{FMNR94}.
Agreement with data was found for the $\Delta \phi$ distribution
after inclusion of a modest intrinsic transverse momentum
$\langle k_T^2 \rangle^{1/2} \sim 0.7$--$1\,$GeV.

The theoretical $p_T(Q\bar{Q})$ distributions, on the other hand,
are too soft unless much larger $k_T$ values are used.
Although it is mostly the intrinsic
$k_T$ of gluons that enters heavy-quark production and not the $k_T$ of
quarks that is known from the Drell--Yan process to be of the order
of $700\,$MeV, it yet appears unreasonable to have a $k_T$ that
is of the order of the hard scale, namely $k_T \sim m_Q$.
Three possible explanations come up.
First higher-order corrections can broaden the $p_T(Q\bar{Q})$ shape.
After all, the $O(\alpha_s^3)$ calculation, although being of NLO
accuracy for the total cross section and single-inclusive distributions,
is still of leading order for the $\Delta \phi$ and $p_T(Q\bar{Q})$ ones.
Second, further non-perturbative contributions besides $k_T$
could be significant. And third, experimental
high-$p_T(c\bar{c})$ data are still rather
limited in statistics and might possibly come down.
New measurements of charm hadroproduction by WA92 are on their way .
The different hypotheses to explain the high-$p_T(Q\bar{Q})$ discrepancy
can also be tested in photoproduction
at HERA where high-$p_T(Q\bar{Q})$ data should soon become available.

\subsection{Bottom production at collider energies}
Owing to its heavier mass, not only the total $b$ cross section but also
differential distributions of the $b$-quark and $b$-flavoured hadrons
can be predicted with higher accuracy (at least, if $m_Q/\sqrt{s}$
is not too small). In fact, the $p_T$ distribution measured
at both $630\,$GeV and $1.8\,$TeV is now found to be consistent (Fig.~4)
with the NLO predictions
\cite{FMNR94}: On the one hand, the experimental data
from the Tevatron `came down'
through improved $b$ tagging, $b\rightarrow J/\psi$, $\psi'$ via
secondary $b$ vertices, $B \rightarrow J/\psi + K^{(*)}$, $b\bar{b}$
correlations, $b \rightarrow \mu X$ identification in D0. On the other hand,
the theoretical prediction `came up', partly because PDF are now known
to steeply increase in the relevant $x$ range \cite{HERA}.

The biggest
change in ref.\ \cite{FMNR94} compared to previous estimates is, however,
that $\Lambda$ is allowed to be as large as the central LEP value
($\Lambda_5 \sim 300\,$MeV or $\Lambda_4 \sim 420\,$MeV). The upper curve
in Fig.~4 is obtained by stretching all parameters to their extremes:
small quark mass ($m_b = 4.5\,$GeV), small renormalization
scale ($\mu_R = \mu_F
= \sqrt{p_T^2 + m_b^2}/2$), and large QCD scale ($\Lambda_5 = 300\,$MeV).
Yet, at the same time the MRSA PDF parametrization \cite{MRSA} with
$\Lambda_5 = 151\,$MeV is being used. First estimates \cite{FMNR94}
indicate that varying $\Lambda$ within a limited range, without
refitting the PDF, is not as large an overestimate of the systematic
effect of the $\Lambda$ uncertainty on the $b$ cross section as one might
expect at first. Given the potential sensitivity of the $b$ cross section
on small-$x$ effects, further theoretical studies should
certainly be pursued.

\section{Bound-state production}
The production of quarkonium states below the open charm/bottom
thresholds presents a particular challenge to QCD. Because of
the relatively large quark masses,
$c$ and $b$ production are perturbatively calculable,
see previous section.
However, the subsequent transition from the
predominantly colour-octet $Q{\bar Q}$ pairs to physical
quarkonium states introduces non-perturbative aspects.
Three models of bound-state formation are discussed in the literature.
In order of increasing sophistication, these are the
colour-evaporation model
\cite{Einhorn,Schuler},
the colour-singlet model
\cite{CS,Schuler},
and the one based on a new factorization formula.

In QCD, the total angular momentum $J$, the parity $P$,
and the charge conjugation $C$ are exactly conserved quantum numbers.
Hence the energy eigenstates $|H\rangle$ of heavy quarkonium
can be labelled by the quantum numbers $J^{PC}$ (besides
a quantum number $n=1,2,\ldots$, to distinguish states identical
apart from their mass $M$ as the $1^{--}$ states $J/\psi$ and $\psi'$).
Phenomenologically very successful is
the non-relativistic potential model, in which $|H\rangle$ is assumed
to be a pure quark--antiquark state $|Q\bar{Q}\rangle$.
Obviously, the $Q\bar{Q}$ pair must then be in a colour-singlet state
$\underline{1}$
and in an angular-momentum state ${}^{2S+1}L_J$ that is consistent with
the quantum numbers $J^{PC}$ of the meson
($P=(-1)^{L+1}$, $C=(-1)^{L+S}$)
\begin{equation}
 {\mathrm{potential}}\;\;{\mathrm{model:}} \quad
|H(nJ^{PC})\rangle =
      |Q\bar{Q}(n {}^{2S+1}L_J,\underline{1})\rangle
\ .
\label{Hpotmodel}
\end{equation}
Here $S=0,1$ is the total  spin of the quark and antiquark,
$L=0,1,2, \ldots$ or $S,P,D,\ldots$ is the orbital angular momentum,
$J$ is the total angular momentum,
and $n$ denotes the principal quantum number.

\subsection{Colour-singlet model}
\label{CSsection}
In the colour-singlet model, the dominant production mechanism of
a heavy quarkonium is assumed to be the one in which the quarkonium
is produced {\em at short distances} in a colour-singlet $Q\bar{Q}$
state with the correct quantum numbers. Hence the
cross section is given by the factorized form ($v=x_F$, $p_T$, $\ldots$):
\begin{equation}
  \frac{d\sigma[H(nJ^{PC})](s,v)}{dv} = F_{nL}\,
     \frac{d\sigma[Q\bar{Q}(n{}^{2S+1}L_J,\underline{1})](s,v)}{dv}
\ .
\label{CScross}
\end{equation}
The non-perturbative probability $F_{nL}$ for the $Q\bar{Q}$ pair to
form the bound state $H$ is given in a calculable way
in terms of the radial wave function or its derivatives
\begin{equation}
  F_{nL} \propto \frac{|R_{nL}^{(L)}(0)|^2}{M_H^{3+2L}}
\label{Fwave}
\end{equation}
and can either be calculated using a phenomenological potential or extracted
from the $H$ decay widths that are given by an expression similar
to (\ref{CScross}) \cite{Schuler}.

Actually, (\ref{CScross}) gives the dominant cross section only if the
relevant momentum scale $Q$ is of the order of the heavy quark mass.
An example is the total hadroproduction cross section
of $J/\psi$ where $Q^2 \sim \hat{s} \sim 4 m^2$.
At large scale $Q$, however, the short-distance production becomes
suppressed by a factor $m^2/Q^2$ with respect to production via
fragmentation that then gives the leading-twist cross section
in $1/Q^2$ and $1/m^2$.
A long-known example \cite{Kuhn81} is $J/\psi$ production
in $Z^0$ decays where $c\bar{c}$ pair production followed by
the fragmentation $c\rightarrow J/\psi$ dominates over $Z^0 \rightarrow
J/\psi gg$.

Similarly, fragmentation processes start to dominate
$p\bar{p} \rightarrow H + X$ at high $p_T$ \cite{Braaten93}.
In this case, important
contributions arise from gluon fragmentation
\begin{eqnarray}
\lefteqn{
 \frac{d\sigma[p\bar{p} \rightarrow H(p)+X](s,p_T)}{dp_T} =
}\nonumber\\ & &
\int_0^1 dz \
   \frac{d\hat{\sigma}[p\bar{p} \rightarrow g(p_T/z) + X]
  (s,p_T,\mu)}{dp_T} \  D_{g \rightarrow H}(z,\mu, m)
\ .
\label{fragcross}
\end{eqnarray}
The fragmentation functions $D_{a \rightarrow H}(z,\mu,m)$ specify the
probability for partons $a$ (gluons, light and heavy quarks)
to hadronize into the hadron $H$ as a function of its longitudinal
momentum fraction $z$ relative to $a$. Large logarithms of $p_T/\mu$ in the
parton cross sections $\hat{\sigma}$ are avoided by
choosing the (arbitrary) factorization scale $\mu$ of the order
of the large scale $p_T$.
Large logarithms of $\mu/m$ then necessarily appear in the fragmentation
functions $D_{a\rightarrow H}(z,\mu,m)$, but they can be summed up
by evolving the ``input" distributions
$D^{(0)}_{a\rightarrow H}(z) \equiv D_{a\rightarrow H}(z,m,m)$
with the standard evolution equations.

In the colour-singlet model, the fragmentation functions
at the input scale $m$, $D^{(0)}_{a\rightarrow H}(z)$
can be calculated as a series in $\alpha_s(m)$
by assuming that they take the same factorized form as
(\ref{CScross}). For example, the lowest-order
diagrams that contribute to gluon fragmentation into $J/\psi$
are $g \rightarrow c\bar{c}({}^3S_1,\underline{1}) g g$, so that
\begin{equation}
  D_{g \rightarrow J/\psi}^{(0)}(z) = \left(\frac{\alpha_s(m)}{m}\right)^3
  |R_{1S}(0)|^2 f(z) + O(\alpha_s^4)
\ ,
\label{gfragpsi}
\end{equation}
where $f(z)$ is a calculable function.
Note that the kinematic regime
in which the $a\rightarrow H$ fragmentation graphs become
important occurs when the lab-frame energy $E_a$ of the parton $a$
is large, but its squared four-momentum $p_a^2$ is close to the
square of the charmonium bound-state's mass $\approx 4 m^2$. Terms
subdominant in the ratio $p_a^2/E_a^2$ are therefore neglected.

In contrast to $S$-wave fragmentation functions \cite{Braaten93,BCY93},
the colour-singlet contributions to fragmentation functions
into $\chi_{Q J}$ state are, however, singular. The process
$g \rightarrow c\bar{c}({}^3P_J,\underline{1}) + g$ diverges
logarithmically \cite{Braaten94} (analogously
$c \rightarrow c\bar{c}({}^3P_J,\underline{1}) + c$ \cite{Chen93})
\begin{equation}
  D_{g\rightarrow \chi_{J}}^{(0)}(z) =
    \frac{\alpha_s^2(m)}{9\pi}\;
  \frac{|R'_P(0)|^2}{m^5}  \left\{
  (2J+1) \left[ \frac{z}{(1-z)_+} + \ln \frac{m}{\epsilon_0}
 \, \delta(1-z)  \right] + r_J(z) \right\}
\ .
\label{naivefragchi}
\end{equation}
Here the infrared divergence, associated with the soft limit of the
final-state gluon has been made explicit by the introduction of a
lower cutoff $\epsilon_0$ on the gluon energy in the quarkonium rest frame.
The presence of the infrared sensitive term clearly spoils the factorization
assumption of the colour-singlet model.

In order to still separate the
long- and short-distance contributions (at least) a second non-perturbative
parameter has to be introduced
\begin{equation}
   D_{g\rightarrow \chi_{J}}^{(0)}(z) = \frac{d_1^{(J)}(z;\lambda)}{m^2}
 \, O_1 + d_8(z)\ O_8(\lambda)
\ .
\label{imfragchi}
\end{equation}
The first term in (\ref{imfragchi}) describes the emission
of a perturbative gluon with energy above some cutoff $\lambda$
\begin{eqnarray}
 O_1 & = & \frac{9}{2\pi}\ |R'_P(0)|^2
\nonumber\\
 d_1^{(J)}(z,\lambda) & = & \frac{2}{81} \frac{\alpha_s^2(m)}{m^3} \left\{
  (2J+1) \left[ \frac{z}{(1-z)_+} + \ln \frac{m}{\lambda}
 \, \delta(1-z) \right] + r_J(z) \right\}
\ .
\label{firstterm}
\end{eqnarray}
In the second term
\begin{eqnarray}
 O_8(\lambda) & = &  \frac{16}{27\pi} \; \alpha_s
  \ln\frac{\lambda}{\epsilon_0}\; \frac{(2J+1)\ H_1}{m^2}
\nonumber\\
d_8(z) & = & \frac{\pi}{24}\; \frac{\alpha_s(m)}{m^3} \, \delta(1-z)
\ ,
\label{secondterm}
\end{eqnarray}
the presence of the infrared scale $\epsilon_0$ indicates that
$O_8$ actually has to be considered as an additional non-perturbative
parameter besides $R'_P(0)$, as expression (\ref{secondterm}) for
$O_8$ can at best give the perturbative part of $O_8$.

Even with improved $P$-wave factorization,
the colour-singlet model still fails in the description of
data on high-$p_T$ prompt charmonium production.
Such data now become available at the Tevatron
where vertex detectors can be used to separate
the charmonium states coming from $b$ quarks from those that are
produced by QCD interactions.
Through the inclusion of fragmentation mechanisms
the theoretical predictions for prompt $J/\psi$ production
can be brought to within a factor of $3$ of the data \cite{BDFM94,Greco94},
close enough that the remaining discrepancy might be attributed
to theoretical uncertainties (Fig.~5).

However, this conclusion of a successful description of $J/\psi$ production
relies on the postulation of a very large
$g \rightarrow \chi_{cJ}(1P)$ fragmentation contribution\footnote{
This large contribution follows from an optimistic choice of $O_8$
in (\ref{imfragchi}).}
where the $\chi_{cJ}$ subsequently decay into $\gamma J/\psi$.
Since such a contribution is absent for $\psi'$
the prediction of its production rate falls a factor of $30$ below the
data (Fig.~5), casting doubts\footnote{The ad hoc introduction of
(not-yet) observed charmonium states above the $D\bar{D}$ threshold
such as higher $P$-wave or $D$-wave states appears quite questionable.
Only with very optimistic production rates and branching fractions
into $\psi'$ can one account for the observed rate for prompt $\psi'$
production \cite{Sridhar94}.}
on whether the $J/\psi$ production is correct at all.
These doubts are strengthened by the observation that
the ratio of $\psi'$ to $J/\psi$ measured at high transverse momenta
at the Tevatron is quite compatible \cite{Brodsky94,Gavai95}
with the $p_T$-integrated fixed-target and ISR data
where direct $J/\psi$ production is known \cite{Schuler}
to dominate the indirect production via $\chi_{cJ}$ decays.
Hence the major part of prompt high-$p_T$ $J/\psi$'s should be directly
produced rather than originate from radiative $\chi_{cJ}$ decays.

The breakdown of the simple factorization into a single long-distance
matrix element and a short-distance Wilson coefficient,
discussed above for gluon fragmentation into $P$-wave quarkonia, has,
in fact, previously been pointed at for the cases of
$B$-meson decays into $\chi_{cJ}$ \cite{Bodwin92},
hadronic $\chi_{QJ}$ decays,
and total cross sections of $\chi_{QJ}$ hadroproduction \cite{Schuler}.
Also these analyses showed that,
even if the colour-singlet model is extended through
the inclusion of colour-octet mechanisms for $P$-wave states
(analogues of the $O_8$ term in (\ref{imfragchi}),
see also section~\ref{newfactsect}), the model
cannot accommodate for all the data on quarkonium production and decays.

It must therefore be concluded that
also in the case of $S$-waves, a generalization of the na\"{\i}ve
factorization \`{a} la (\ref{CScross},\ref{gfragpsi}) is necessary, at least
in cases where the colour-singlet mechanism is suppressed by a
short-distance coefficient, i.e.\ by a high power of $\alpha_s(m)$,
cf.\ (\ref{gfragpsi}).
A factorization of long- and short-distance physics more generalized
than the one assumed in the colour-singlet model
will be presented in section~\ref{newfactsect}. First, however,
a quarkonium production-model diametrally opposite to the
colour-singlet model is discussed.

\subsection{Colour-evaporation model}
The starting, i.e.\ perturbatively calculated cross section
for heavy quarkonium production
in the colour-evaporation model is the (usual) cross section of
open heavy-quark pair production that is
summed over all spin and colour states. Hence all the information on the
non-perturbative transition of the $Q\bar{Q}$ pair to
the heavy quarkonium $H$ of quantum numbers $J^{PC}$
is contained in ``fudge factors" $F_{nJ^{PC}}$ that a priori may depend
on all quantum numbers
\begin{equation}
  \frac{d\sigma[H(nJ^{PC})](s,v)}{dv} = F_{nJ^{PC}}\,
     \frac{d\tilde{\sigma}[Q\bar{Q}](s,v)}{dv}
\ .
\label{CEcross}
\end{equation}
In (\ref{CEcross}) $\tilde{\sigma}[Q\bar{Q}]$ is
the total ``hidden"  cross section of (open) heavy-quark production
calculated by  integrating over the $c\bar{c}$ pair mass
(in the case of charmonium) from $2m_c$ to $2m_D$.
For example, in hadronic collisions
at high energy, the dominant production mechanism is gluon fusion,
so that (cf.\ (\ref{ABcross})):
\begin{eqnarray}
\lefteqn{
 \tilde{\sigma}[c\bar{c}](s) =
 \int_{4m_c^2}^{4m_D^2} d\hat s \int dx_1 dx_2
}\nonumber\\ & &
  f_{g/A}(x_1,m^2)
 f_{g/B}(x_2,m^2) \widehat \sigma[gg\rightarrow c\bar{c}X](\hat s)
    \delta(\hat s - x_1x_2s)
\ .
\label{gluonfusioncross}
\end{eqnarray}
Note that $\tilde\sigma[c\bar{c}]$ is the spin-summed cross section
and that the heavy-quark pair can be both in a colour-singlet
and a colour-octet state
(in LO, $q\bar{q}$ annihilation produces only colour-octet
$c\bar{c}$ pairs, while $gg$ fusion also leads to colour-singlet states).
The colour-octet cross section is the dominant one.
The $\c$ configuration arranges itself into a definite outgoing
charmonium state
by interacting
with the collision-induced colour field (``colour evaporation").
During this process, the $c$ and the $\bar c$ either combine with
light quarks to produce charmed mesons, or they bind
with each other to form a charmonium state.

As shown in ref.\ \cite{Gavai95} more than half of
the subthreshold cross section \S~in fact goes into
open charm production (assuming $m_c < m[\eta_c]/2 \sim 1.5\,$GeV);
the additional energy needed to produce charmed hadrons
is obtained (in general non-perturbatively) from the
colour field in the interaction region.
The yield of all charmonium states below the $D{\bar D}$ threshold is
thus only a part of the total sub-threshold cross section: in this
aspect the modern version of the model \cite{Gavai95}
is a generalization of the original colour-evaporation model
\cite{Einhorn,Schuler}, which neglected the contribution of \S~to
open charm production. Using duality arguments, it equated \S~to the
sum over the charmonium states below the $D{\bar D}$ threshold.

Neither the division of \S~into open charm and charmonia nor the
relative charmonium production rates are specified by the generalized
colour-evaporation model. Hence its essential prediction is that the
dynamics of charmonium production is that of $\s$, i.e.\ the
energy dependence, $x_F$- and $p_T$-distributions of $H$ are identical
to those of the free $c\bar{c}$ pair.
In particular, ratios of different charmonium
production cross sections should be energy-, $x_F$-, and $p_T$-independent.
In other words, the non-perturbative factors $F_{nJ^{PC}}$ should
be universal constants whose values may, however, depend on the
heavy-quark mass. In contrast to earlier expectations \cite{Schuler},
a recent comprehensive comparison of the generalized colour-evaporation
model with data indeed confirmed these expectations \cite{Gavai95}.

Figure~6 shows the ratio of \J~production from the decay $\chi_c
\to \gamma~\j$ to the total \J~production rate, which
provides a measure of the $\chi_c/(\j)$ rate, and Fig.~7 shows
the measured \P/(\J) ratio.
Both ratios are found to be independent of the incident energy,
the projectile (pion or proton), and the target (from
protons to the heaviest nuclei).
Moreover, it is noteworthy \cite{Brodsky94,Gavai95}
that the ratio \P/(\J) measured at
high transverse momenta at the Tevatron
is quite compatible with the $p_T$-integrated fixed-target and ISR data
(Fig.\ 8). Also the available bottomonium data
agree with constant production ratios \cite{Gavai95}.

Figures~9 and~10 show the energy dependence of \J~and $\Upsilon$ production
in hadronic collisions; the agreement with data over a wide range
is rather impressive. Because the data generally give the sum
of \U, $\u'$ and $\u''$ production,
the measured cross section
for the sum of the three \U~states in the dilepton decay
channel is shown, denoted by $B(d\sigma/dy)_{y=0}$.
Using the average values of the $\u''$~to \U~and $\u'$~to
\U~production ratios, $0.53\pm 0.13$ and $0.17\pm 0.06$, respectively,
fits \cite{Gavai95} to the data in Figs.~9 and~10 yield
\begin{equation}
  F_{11^{--}} \approx \left\{ \begin{array}{ll}
                       2.5\times 10^{-2} & \mathrm{charm}\\
                       4.6\times 10^{-2} & \mathrm{bottom}
                       \end{array} \right.
\ .
\label{Fnlsvalues}
\end{equation}

The hidden heavy-flavour cross sections $\tilde\sigma[Q\bar{Q}]$ in Figs.~9
and~10 were calculated in NLO using the MRS D$-'$ parametrization \cite{D0}
of PDF with
$\mu_F = \mu_R = 2 m_c = 2.4\,$GeV and $\mu_F = \mu_R=m_b = 4.75\,$GeV,
respectively.
These parameters provide an adequate description of open heavy-flavour
production \cite{Mcgaughey}, cf.\ section~1.
Results similar to (\ref{Fnlsvalues}) are obtained if one uses
other choices of the parameters that are tuned to the open heavy-flavour
data, for instance the GRV parametrization \cite{GRV} with
$\mu_F = \mu_R = m_Q$, $m_c = 1.3\,$GeV, and $m_b = 4.75\,$GeV.
{}From (\ref{Fnlsvalues}) one concludes that
the fraction of \S~producing charmonium rather than open charm
is about 10\%.

Equally good agreement is found
for the energy dependence of \J~production with pion beams. However,
the fraction of \J~in the hidden charm cross section must
be slightly higher to reproduce the pion
data well, with $F_{11^{--}} =
0.034$ for a good fit. This may well be due to
greater uncertainties in the pionic parton distribution functions.
Consistency with the colour-evaporation model is also found  for
the longitudinal momentum dependence of charmonium production.
The calculations for the $x_F$ dependence of \J~ production
agree with data (Fig.~11) from low-energy $\bar{p}p$ interactions (where
$q\bar{q}$ annihilation is important) up to $pp$ interactions
at the highest available energy and out to the largest $x_F$ values
probed (where the ``intrinsic charm" mechanism could have been important).

Concerning the $p_T$ distribution, the model
provides essentially no prediction for low-$p_T$ charmonium production.
There is the intrinsic
transverse momentum of the initial partons, the
intrinsic momentum fluctuations of the colour field which neutralizes
the colour of the $\c$ system in the evaporation process
and, at larger $p_T$, higher-order perturbative terms. Since there is no
way to separate these different contributions in the low-$p_T$ region,
the model has no predictive power.

On the other hand, the high-$p_T$ tail should be successfully
describable in the colour-evaporation model,
presumably with the same normalization (\ref{Fnlsvalues}).
The time scale to form a quarkonium bound state is much larger
than the one to produce the (compact) $c\bar{c}$ pair. Hence the
fraction of the $c\bar{c}$ cross section (in $2m_c < M_{c\bar{c}} < 2m_D$)
that becomes a $J/\psi$ (or $\psi'$) should be independent of the
$c\bar{c}$ production, i.e.\ the same for low-$p_T$ and high-$p_T$
processes. By the same argument, the universal ratio of
$\psi'$ to $J/\psi$ production, observed experimentally, can
depend only on the relative magnitude of the respective
wave functions at the origin
\begin{eqnarray}
 \frac{\Gamma(\psi'\rightarrow e^+e^-)}
      {\Gamma(J/\psi\rightarrow e^+e^-)}
 \left({M_{J/\psi}\over M_{\psi'}} \right)^3 & = &
  {\sigma(\psi') \over \sigma_{dir}(J/\psi)}
\label{psiprime}
\\ \nonumber
 & = & \left[{1\over
1-\sigma(\chi_c\rightarrow J/\psi)/\sigma(\j)}\right]
  \left[
{\sigma(\psi') \over \sigma(\j)}\right]_{\rm exp}
\ .
\end{eqnarray}
Relation (\ref{psiprime}) holds to very good approximation
\cite{Brodsky94,Gavai95}. Note, finally, that gluon fragmentation
$gg \rightarrow g^\star g$ with $g^\star \rightarrow c\bar{c}
\rightarrow J/\psi$ is part of the lowest-order ($O(\alpha_s^3)$)
diagrams describing high-$p_T$ charmonium production in hadronic
collisions, while charm fragmentation $gg\rightarrow c^\star\bar{c}$ with
$c^\star\rightarrow J/\psi$ first occurs at $O(\alpha_s^4)$.

\subsection{A new factorization approach}
\label{newfactsect}
The two previous sections discussed two extreme scenarios
to describe production cross sections (and decay rates) of
heavy quarkonia. In the colour-evaporation model (\ref{CEcross}),
no constraints are imposed on the colour and angular momentum states of
the $Q\bar{Q}$ pair. Non-perturbative QCD effects, mediating the
transition to the colour-singlet bound state $H(J^{PC})$ containing the
$Q\bar{Q}$ pair are assumed to be first universal and secondly
negligible for the dynamics of $H$ ($\sqrt{s}$, $p_T$, etc.\ dependence).
The normalization factors for the various states are not
predictable, but once fixed phenomenologically,
the model is (surprisingly) successful.

The factorization assumption (\ref{CScross})
of the colour-singlet model, on the other hand, says that all
non-perturbative effects are contained in a single term that can
be expressed as the non-relativistic wave function of the
bound state. In turn, relative production rates of different
quarkonium states can be predicted. Moreover, different states may
have different dynamical dependences since only specific short-distance
cross sections contribute to each state. However, the colour-singlet
model fails in two respects. First, predictions for $S$-wave states often
are way off, and second logarithmic infrared divergences spoil the
factorization in the case of $P$-waves, cf.\ section~\ref{CSsection}.

This failure of the colour-singlet model can be traced back to that
of the underlying quark potential model. Relativistic corrections are
essential for a description that is both consistent for $P$-wave states
and successful for $S$-wave states. Recently, a rigorous QCD
analysis of the annihilation decays of heavy quarkonium has been presented
based on recasting the analysis in terms of HQ$\bar{\mathrm{Q}}$ET,
an effective field theory designed precisely for this purpose \cite{Mannel}.
It allows the separation of long and short distances, where the short-distance
contribution may be evaluated perturbatively, i.e.\ as series
in $\alpha_s(m)$. The long-distance part is parametrized in terms
of matrix elements, which are organized into a hierarchy according to
their scaling with $1/m$.

A similar analysis has been performed in the context of non-relativistic
quantum chromodynamics (NRQCD) \cite{NRQCD}. In this approach, the
calculations are organized in powers of $v$, the average velocity of the
heavy (anti-)quark in the meson rest frame. Contributions of different
orders in $v$ are separated according to the ``velocity-scaling" rules.
This factorization formalism has been extended to the production
cross section of heavy quarkonium in processes involving momentum
transfers of order $m$ or larger.

The inclusive production cross section of a quarkonium state $H$ and
the parton fragmentation function into $H$ take the form:
\begin{eqnarray}
\sigma(H) & = &
  \sum_{c=1,8} \sum_{d=6,8,\ldots} \sum_{X}
  \frac{F_c^{(d)}(X;\lambda)}{m^{d-4}}
  \langle 0 | {\cal O}^H_c(d,X;\lambda) | 0 \rangle
\nonumber\\
D^{(0)}_{a \rightarrow H}(z) & = &
   \sum_{c=1,8} \sum_{d=6,8,\ldots} \sum_{X}
\frac{d_c^{(d)}(X;z,\lambda)}{m^{d-6}}
  \langle 0 | {\cal O}^H_c(d,X;\lambda) | 0 \rangle
\ .
\label{fragfact}
\end{eqnarray}
Equation (\ref{fragfact})
expresses the cross section (and analogously the fragmentation function)
as a sum of terms, each of which factors into a short-distance coefficient
$F_n(\lambda)$
and a long-distance matrix element
$\langle 0 | {\cal O}_n^H(\lambda) \rangle$ (where $n=\{c,d,X\}$).
The coefficients $F_n$ are proportional to the rates for
the production of on-shell heavy quarks and antiquarks from initial-state
gluons and light quarks, and they can be computed as perturbation series
in $\alpha_s(m)$.
They depend on all
the kinematical variables of the production process.
The matrix elements $\langle 0 | {\cal O}_n^H |0 \rangle$
give the probability for the formation of the quarkonium state $H$
from the $Q\bar{Q}$ pair of state $n$, and can be evaluated
non-perturbatively using, for example, QCD sum rules or lattice
simulations. The dependence on the arbitrary factorization scale $\lambda$
in (\ref{fragfact}) cancels between the coefficients and the operators.

The expansion (\ref{fragfact}) is organized into an expansion in powers
of $v^2$. Only a finite number of operators contribute to any given order
in $v^2$. Since the coefficients $F_n$ are calculated as perturbation
series in $\alpha_s(m)$, eq.~(\ref{fragfact}) is really a double expansion
in $\alpha_s(m)$ and $v^2$. For heavy quarkonia, the two expansion
parameters are not independent: $v \approx \alpha_s(m v) > \alpha_s(m)$.
Hence corrections of order $v^n$ must not be neglected compared
to those of order $\alpha_s^n(m)$. Which terms in (\ref{fragfact})
actually contribute to the production of a quarkonium $|H\rangle$
depends on both the $\alpha_s(m)$ expansion of $F_n$ and the
$v^2$ expansion of the matrix elements. The latter is determined by
the variables that specify a given operator ${\cal O}^H_c(d,X)$.
These are the dimension $d$ of the operator, and two further variables
$c$ and $X$, which specify the quantum numbers of the heavy $Q\bar{Q}$ pair
in the Fock-state expansion of the heavy quarkonium $H$.

The state of the $Q\bar{Q}$ pair in a general Fock state
can be labelled by the colour state of the pair, singlet ($c=1$)
or octet ($c=8$), and the angular momentum state ${}^{2S+1}L_J$ of the
pair denoted by $X$. The leading term of the Fock-state expansion
is the pure $Q\bar{Q}$ state (\ref{Hpotmodel}) of the potential model.
Higher Fock states are suppressed by powers of $v$, which follow from the
velocity-scaling rules \cite{NRQCD}.
For example, if the $Q\bar{Q}$ in the dominant
Fock state $|Q\bar{Q}\rangle$ has angular-momentum quantum numbers
${}^{2S+1}L_J$, then the Fock state $|Q\bar{Q}g\rangle$ has an amplitude of
order $v$ only if the $Q\bar{Q}$ pair has total spin $S$ and orbital
angular momentum $L+1$ or $L-1$ (E1 transition). Necessarily, the
$Q\bar{Q}$ pair must be in a colour-octet state. The general
Fock-state expansion therefore starts as
\begin{eqnarray}
  |H(nJ^{PC})\rangle & = & \quad
      O(1)\, |Q\bar{Q}({}^{2S+1}L_J,\underline{1})\rangle
\nonumber\\
 & & +~ O(v) \;\ |Q\bar{Q}({}^{2S+1}(L\pm 1)_{J'},\underline{8})\ g\rangle
\nonumber\\
 & & +~ O(v^2) \ |Q\bar{Q}({}^{2S+1}L_{J},\underline{8})\ gg\rangle
   + \ldots
\nonumber\\
 & & +~ \ldots
\ .
\label{Fockexpand}
\end{eqnarray}

In the case of $J/\psi$ production, for example, one finds up to
and including the
\clearpage
order $v^2$ \cite{NRQCD}:
\begin{eqnarray}
\sigma(J/\psi) & = & \frac{F_1^{(6)}({}^3S_1;\lambda)}{m^2}
  \langle 0| {\cal O}_1^{J/\psi}(6,{}^3S_1;\lambda) |0 \rangle
\nonumber\\
 & & \;\;
 +~  \frac{F_1^{(8)}({}^3S_1;\lambda)}{m^4}
  \langle 0| {\cal O}_1^{J/\psi}(8,{}^3S_1;\lambda) |0 \rangle
 + O(v^4)
\ .
\label{v4jpsi}
\end{eqnarray}
Upon dropping the $O(v^2)$ contribution and identifying
$ \langle 0| {\cal O}_1^{J/\psi}(6,{}^3S_1;\lambda) |0 \rangle =
3 N_C |R_{1S}(0)|^2 /(2\pi)$, eq.\ (\ref{v4jpsi}) reduces to the familiar
factorization formula (\ref{CScross}) of the colour-singlet model.
Generally, the standard factorization formulas of the
colour-singlet model,
which contain a single non-perturbative parameter, are recovered
in the case of $S$-waves at leading order in $v^2$.

However, the factorization formula is the sum of two terms
in the case of $P$-waves.
In addition to the conventional term, which takes into account the
production (or annihilation) of the $Q\bar{Q}$ pair from a colour-singlet
$P$-wave state, there is a second term that involves production
(annihilation) from a colour-octet $S$-wave state. For example,
the gluon fragmentation function into $\chi_{cJ}$ states ($J=0,1,2$)
is \cite{Braaten94}, cf.\ (\ref{naivefragchi},\ref{imfragchi}):
\begin{eqnarray}
 D_{g\rightarrow \chi_{cJ}}^{(0)}(z) & = &
   \frac{d_1^{(8)}(z;{}^3P_J;\lambda)}{m^2}
  \langle 0| {\cal O}_1^{\chi_{cJ}}(8,{}^3P_J;\lambda) |0 \rangle
\nonumber\\ & & \;\;
 +~  d_8^{(6)}(z;{}^3S_1;\lambda)
  \langle 0| {\cal O}_8^{\chi_{cJ}}(6,{}^3S_1;\lambda) |0 \rangle
 + O(v^2)
\ .
\label{v2chi}
\end{eqnarray}

The $J^{++}$ state $|\chi_{cJ}\rangle$ consists predominantly of the
Fock state $|Q\bar{Q}\rangle$, with the $Q\bar{Q}$ pair in a
colour-singlet ${}^3P_J$ state. It also has an amplitude of order $v$
for the Fock state $|Q\bar{Q}g\rangle$, with the $Q\bar{Q}$ pair in a
colour-octet ${}^3S_1$, ${}^3D_1$, ${}^3D_2$, or ${}^3D_3$ state:
\begin{equation}
 |\chi_{Q J} \rangle = O(1) | Q\bar{Q}({}^3P_J,\underline{1}) \rangle
    +  O(v) | Q\bar{Q}({}^3S_1,\underline{8}) g \rangle
    +  O(v) | Q\bar{Q}({}^3D_{J'},\underline{8}) g \rangle + O(v^2)
\ .
\label{chiexpand}
\end{equation}
The Fock state $|Q\bar{Q}\rangle$ contributes to the production at
leading order in $v^2$ through the dimension-$8$ operator
${\cal O}_1^{\chi_{cJ}}(8,{}^3P_J)$. The Fock state $|Q\bar{Q}g\rangle$,
with the $Q\bar{Q}$ pair in a colour-octet ${}^3S_1$ state also contributes
to the production at the same order in $v^2$, through the dimension-$6$
operator ${\cal O}_8^{\chi_{cJ}}(6,{}^3S_1)$, because the latter
scales as $v^{-2}$ relative to ${\cal O}_1^{\chi_{cJ}}(8,{}^3P_J)$.

Equation (\ref{v2chi}), together with the evolution equations for the
matrix elements contain the solution to the problem of the
infrared divergence encountered in (\ref{naivefragchi}).
At leading order in $\alpha_s(m)$,
the coefficient $d_1^{(8)}(z;{}^3P_J;\lambda)$ in (\ref{v2chi})
($\equiv d_1^{(J)}(z,\lambda)$ in (\ref{imfragchi},\ref{firstterm}))
depends logarithmically on the factorization scale $\lambda$, while
$d_8^{(6)}(z;{}^3S_1;\lambda) \equiv d_8(z)$
in (\ref{imfragchi},\ref{secondterm})  is independent of $\lambda$.
In these coefficients, $\lambda$ plays the role of an infrared
cutoff. The $\lambda$-dependence of the short-distance coefficients
cancels the $\lambda$-dependence of the long-distance matrix elements,
for which $\lambda$ plays the role of an ultraviolet cutoff.

At leading order in $v^2$ and $\alpha_s(m)$, the dimension-8
matrix element
$O_1(\lambda) \equiv \
\langle 0 | {\cal O}^{\chi_J}_1(8,{}^3P_J;\lambda) | 0 \rangle$
is renormalization-scale-invariant, while the dimension-8 matrix element
$O_8(\lambda) \equiv
\langle 0 | {\cal O}^{\chi_J}_8(6,{}^3S_1;\lambda) | 0 \rangle$
has a non-trivial scaling behaviour \cite{NRQCD}:
\begin{eqnarray}
 \lambda \frac{d}{d\lambda} O_1(\lambda) & = & 0
\nonumber\\
 \lambda \frac{d}{d\lambda} O_8(\lambda)  & = & \frac{16}{27\pi}
  \alpha_s(\lambda)\,  \frac{(2J+1) O_1}{m^2}
\label{O8evol}
\end{eqnarray}
With the help of (\ref{O8evol}) and (\ref{firstterm},\ref{secondterm})
it is straightforward to show that
\begin{equation}
\lambda \frac{d D^{(0)}_{g\rightarrow \chi_{cJ}}}{d\lambda} = 0
\ .
\label{facttest}
\end{equation}
This is in accordance with the general expectation that physical
quantities, such as fragmentation functions, are renormalization-group
invariants, i.e.\ independent of the arbitrary factorization scale $\lambda$.
(In practice, yet, it might not always be easy to know at which value
of $\lambda$ the matrix elements are evaluated.)

To leading order in $\alpha_s$ the evolution equations (\ref{O8evol})
can be solved analytically with the result
\begin{equation}
 O_8(m)  =  O_8(\lambda) + \frac{16}{27\beta_0} \;
  \ln\frac{\alpha_s(\lambda)}{\alpha_s(m)}\, \frac{(2J+1) O_1}{m^2}
\ .
\label{O8solution}
\end{equation}
This solution may be used to provide an order-of-magnitude estimate
of the colout-octet matrix element $O_8$ in terms of the colour-singlet
matrix element $O_1$ by assuming that $O_8(\lambda)$ can be neglected
compared to the second term in (\ref{O8solution}) for $\alpha_s(\lambda)=1$.
Recall that, to leading order in $v^2$, $O_1 = 9 |R'_P(0)|^2/2\pi$.
Note also that, solving (\ref{O8evol}) for constant $\alpha_s$,
one recovers the approximate solution (\ref{secondterm}).

The new factorization approach provides a consistent framework
for the calculation of $P$-wave production and decays;
$\chi_{cJ}$ decays can well be described and a reasonable value
of $\alpha_s(m)$ is found \cite{Schuler}. Detailed
comparisons of $\chi_{cJ}$ production with experiments have not yet
been performed. Based on (\ref{v2chi}) or (\ref{imfragchi}) one expects
$\chi_{cJ}$ production rates to be approximately proportional to $(2J+1)$.
Preliminary CDF data \cite{private} on the $\chi_{c1}$ to $\chi_{c2}$ ratio
at high $p_T$, where gluon fragmentation is the dominant
production mechanism, indeed seem to confirm this expectation.

The comparison is less fortunate for the total production rates
measured at fixed-target energies. Experimentally, the ratio
of $\chi_{c1}$ to $\chi_{c2}$ production is about
$2:3$ \cite{Schuler}. This still seems to hold
at $\sqrt{s}$ sufficiently large
for the gluon fusion to dominate over quark-initiated processes.
However, one then expects, theoretically,
$\sigma(\chi_{c1})/\sigma(\chi_{c2}) = 0$
in leading order in $v^2$ and $\alpha_s(m)$:
The Landau--Yan theorem forbids $\chi_{c1}$
production through the fusion of two gluons, hence the
colour-singlet matrix element
$\langle 0| {\cal O}^{\chi_{c1}}_1(8,{}^3P_1) |0\rangle$ does not
contribute at $O(\alpha_s^2)$. However, there is no contribution
either from the colour-octet matrix element
$\langle 0| {\cal O}^{\chi_{c1}}_8(6,{}^3S_1) |0\rangle$: two-gluon fusion
into a coloured ${}^3S_1$ state is absent, since in the non-relativistic
limit the corresponding amplitude is equal to zero.
Hence corrections to the ratio are truly of $O(\alpha_s(m))$
or of $O(v^2)$, i.e.\ there is no $O(v^2)$ correction that is
enhanced by a short-distance factor $1/\alpha_s(m)$.
(Such a correction does, however, exist in the case of $q\bar{q}$
annihilation.)

Similar problems exist in the description of $J/\psi$ (and $\psi'$,
collectively denoted by $\psi$ in the following) production and decays.
Up to and including the $O(v^2)$ its Fock-state expansion is
\begin{eqnarray}
 |\psi(1^{--}) \rangle & = &
    \quad O(1) | Q\bar{Q}({}^3S_1,\underline{1}) \rangle
    +~  O(v) | Q\bar{Q}({}^3P_J,\underline{8}) g \rangle
\nonumber\\ & & + \sum_{c=1,8} \left \{
   O(v^2) | Q\bar{Q}({}^3S_{1},\underline{c}) gg \rangle
   +  O(v^2) | Q\bar{Q}({}^3D_{J},\underline{c}) gg \rangle
  \right\}
\nonumber\\ & &
   +\;  O(v^2) | Q\bar{Q}({}^1S_{0},\underline{8}) g \rangle
 + O(v^3)
\ .
\label{psiexpand}
\end{eqnarray}
Hence corrections to the predictions of the colour-singlet model
are truly of $O(v^2)$, i.e.\ to $O(1)$ the new factorization approach
coincides with the colour-singlet model that is known to be way-off
the data. Large relativistic corrections are therefore needed.

In the context of a meaningful expansion in powers of $v^2$,
large relativistic corrections can only be accommodated if contributions,
suppressed by power(s) of $v^2$ become important because the leading
contribution is suppressed by additional power(s) of $\alpha_s(m)$.
However, the $O(v^2)$ corrections
are not enhanced by a short-distance factor $1/\alpha_s(m)$ since they
are genuine $O(v^2)$ corrections to the production of a
colour-singlet ${}^3S_1$ state, cf.\ (\ref{v4jpsi}). Only at
$O(v^4)$ does such an enhancement occur: a factor $1/\alpha_s(m)$
in gluon--gluon fusion to $J/\psi$ via
$\langle 0| {\cal O}_8^{\psi}(6,{}^1S_0) | 0\rangle$ and
$\langle 0| {\cal O}_8^{\psi}(8,{}^1P_{0,2}) | 0\rangle$
(relevant for $\psi$ production at fixed-target energies), and
a factor $1/\alpha_s^2(m)$ in gluon fragmentation into $\psi$ via
$\langle 0| {\cal O}_8^{\psi}(6,{}^3S_1) | 0\rangle$. Treating
the latter matrix element as a free parameter, the
description of high-$p_T$ $\psi'$ production at the
Tevatron can indeed be rescued \cite{Fleming94},
but clearly more work is needed before the new factorization formalism
is established as (the) successful theory of quarkonium production.

Further work can come from three sources. First, further phenomenological
studies are needed. Ideally, one would like to have a global analysis
of the data on charmonium production from all high-energy processes.
Comparisons with bottomonium production should confirm that relativistic
corrections indeed decrease as the heavy-quark mass increases.

Secondly, more theoretical analyses of the foundation of the factorization
formalism of NRQCD \cite{NRQCD} in (full) QCD are needed.
A key ingredient is the velocity-scaling rules,
both for the ordering of operators of a given dimension and the
Fock-state expansion. Using perturbation theory, each additional gluon
associated with the (assumed) dominant $Q\bar{Q}$ pair is ascribed
an extra power of $v$ via the identification $v \sim \alpha_s(mv)$, valid
for a colour-Coulomb potential. Although this estimate may be underlined
by the multipole expansion, we do not know of any rigorous derivation.
One way would be the extension of the factorization of
HQ$\overline{\mathrm{Q}}$ET \cite{Mannel} from quarkonium decays to their
production.
Additionally, implications of spin symmetry \cite{Cho}
for the production (and decays)
of heavy quarkonia should be investigated further.

Thirdly, more experimental data will improve our understanding
of bound states of heavy quarks. Concerning their production,
one would in particular like to see ratios
of the production rates of various charmonium states; as a function of $p_T$
at the Tevatron and HERA, and their $x_F$ dependence in hadro- and
photoproduction. Interesting would also be the observation of
a spin alignment of the heavy quarkonium states. Last but not least,
valuable information will come from  charmonium production
in $e^+e^-$ collisions through improved measurements of $b$-decays
into charmonium, observation of a fragmentation contribution, and (at
LEP2) charmonium production in two-photon collisions.\hfill\\[2ex]

\noindent
{\bf Acknowledgements}\hfill\\[1ex]
I am grateful to M.\ Mangano, G.\ Ridolfi, and R.\ Vogt
for providing me with figures from their works.

\clearpage
\begin{figure}
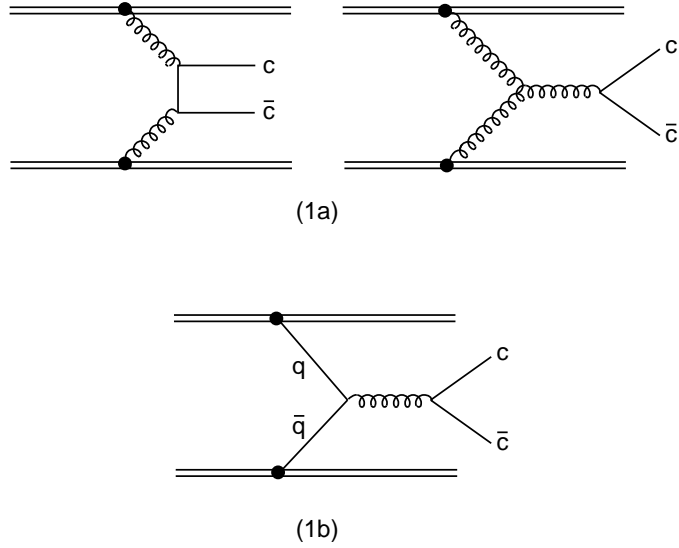

\caption{
Lowest-order contributions to heavy-quark hadroproduction.
}
\end{figure}
\begin{figure}
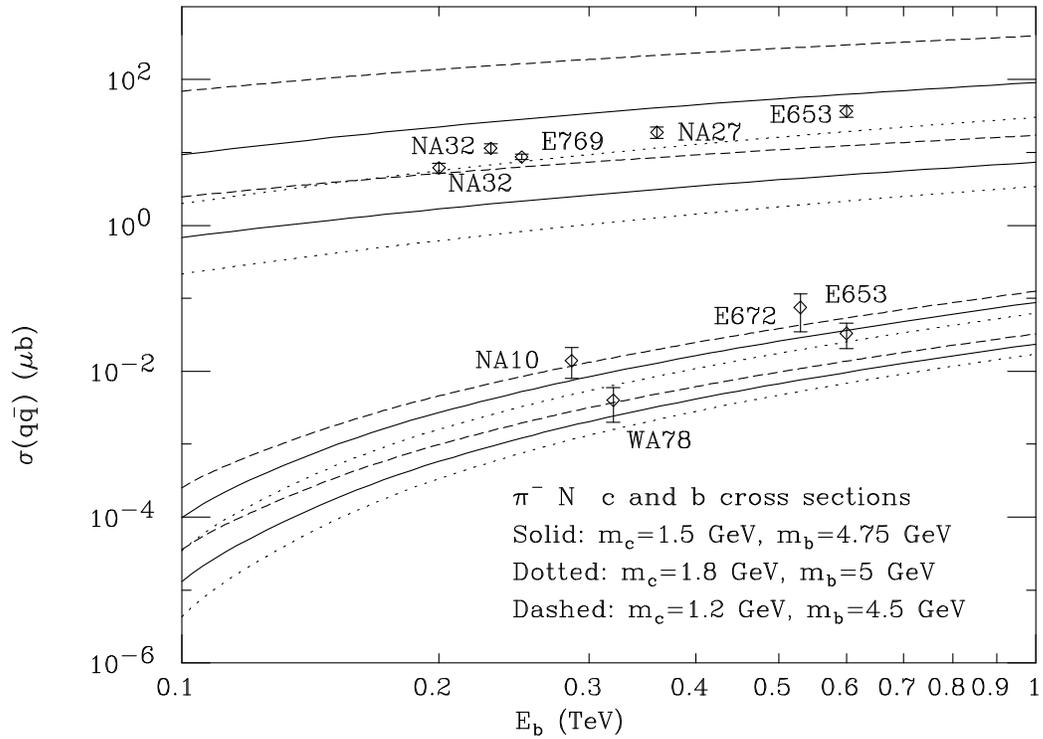

\caption{
 Cross sections for $b$ and $c$ production in $\pi N$ collisions.
 For details see text
 (from ref.~\protect\cite{FMNR94}).
}
\end{figure}
\begin{figure}
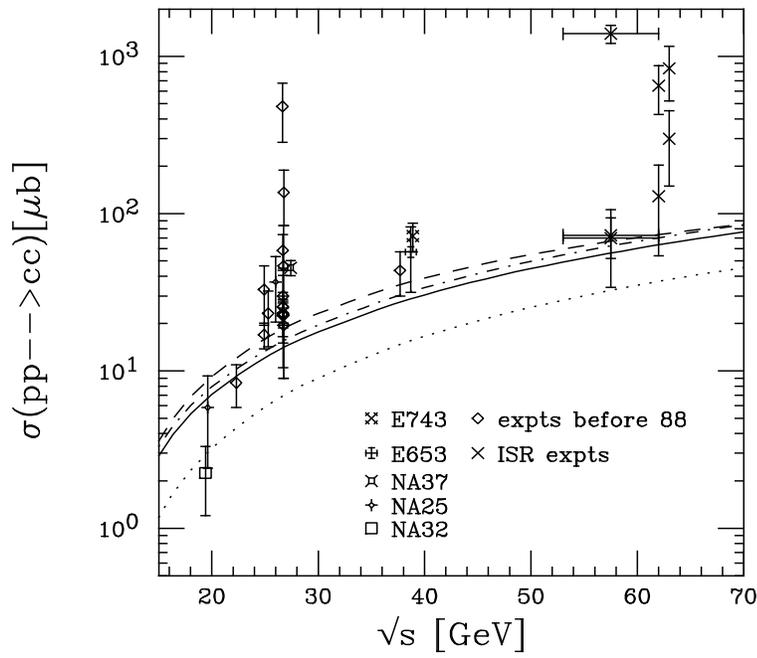

\caption{
 Total charm production cross sections from $pp$ and $pA$ measurements
 compared to calculations.  The
 curves are:  MRS D$-'$ $m_c=1.2$ GeV, $\mu = 2m_c$ (solid); MRS
 D0$^\prime$ $m_c=1.2$ GeV, $\mu = 2m_c$ (dashed); GRV HO $m_c = 1.3$ GeV,
 $\mu = m_c$ (dot-dashed); GRV HO $m_c = 1.5$ GeV, $\mu = m_c$ (dotted)
 (from ref.~\protect\cite{Mcgaughey}).
}
\end{figure}
\begin{figure}
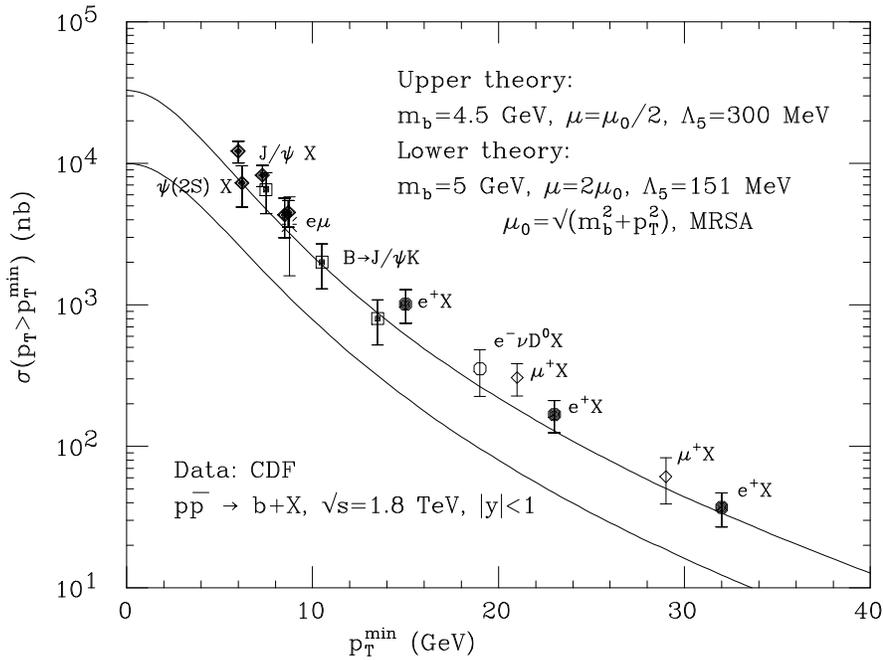

\caption{
 The $b$-quark cross sections at CDF.
 For details see text
 (from ref.~\protect\cite{FMNR94}).
}
\end{figure}
\begin{figure}
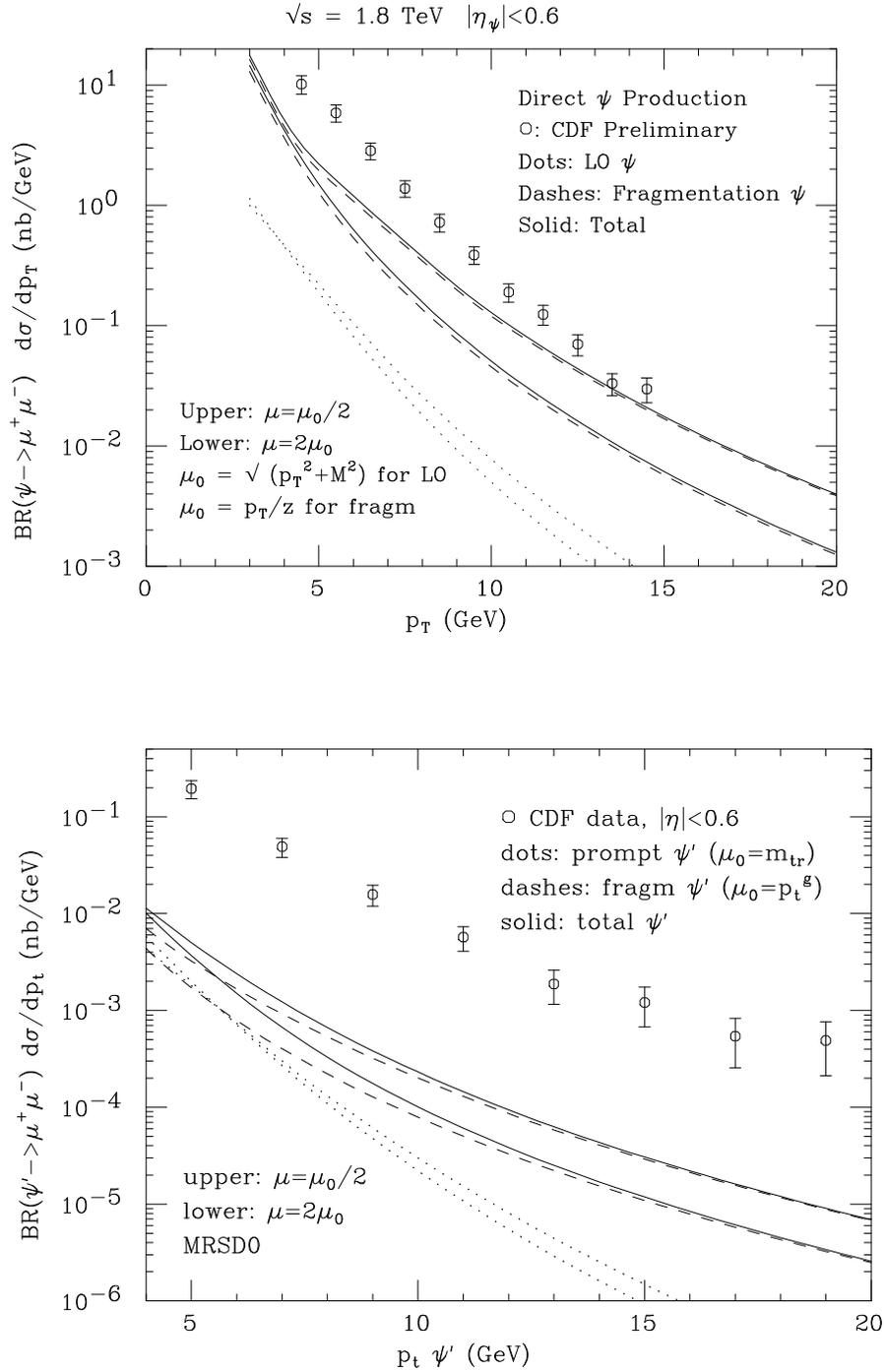

\caption{Preliminary CDF data for prompt $J/\psi$ and $\psi'$ production
 compared with theoretical predictions of the total fragmentation
 contribution (solid curves) and the total leading-order contribution
 (dashed curves)
(from ref.~\protect\cite{BDFM94}).
}
\end{figure}
\begin{figure}
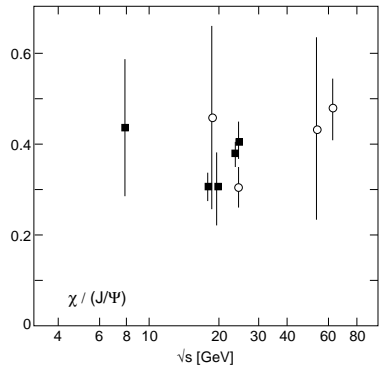

\caption{
 The ratio of $(\chi_{c1} + \chi_{c 2}) \rightarrow J/\psi$
 to total $J/\psi$ production as a function of
 c.m.\ energy $\protect\sqrt{s}$,
 by proton (open symbols) and pion beams (solid symbols)
 (from ref.~\protect\cite{Gavai95}).
}
\end{figure}
\begin{figure}
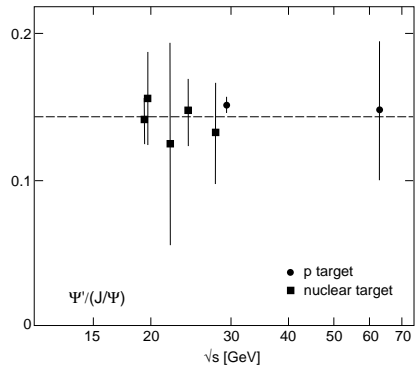

\caption{
 The ratio of $\psi'$ to $J/\psi$ production as a function of c.m.\
 energy $\protect\sqrt{s}$, on proton (circles) and nuclear targets
 (squares) (from ref.~\protect\cite{Gavai95}).
}
\end{figure}
\begin{figure}
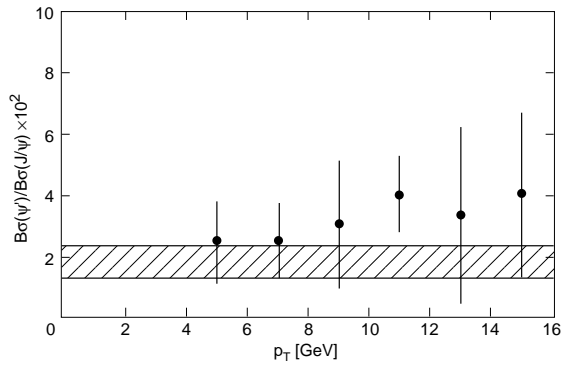

\caption{
 The ratio of $\psi'$ to $J/\psi$ production as a function of transverse
 momentum; the shaded strip shows the average value of Fig.~7
 (from ref.~\protect\cite{Gavai95}).
}
\end{figure}
\begin{figure}
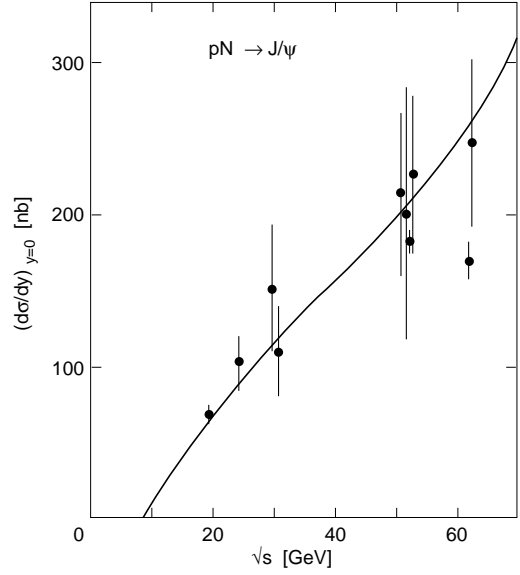

\caption{
  The differential \J~production cross section
  $(d\sigma[pN \to J/\psi X]/dy)=
  2.5\times10^{-2}~(d\tilde\sigma[c\bar{c}]/dy)$ at $y=0$, calculated with
  MRS D$-'$ PDF, compared with data
 (from ref.~\protect\cite{Gavai95}).
}
\end{figure}
\begin{figure}
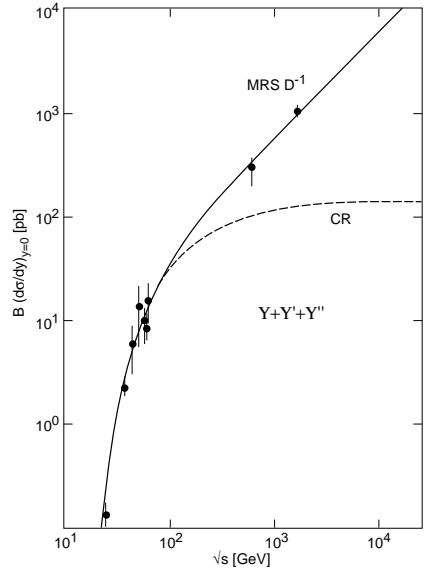

\caption{
 Energy dependence of $\Upsilon$ production in $pN$ collisions
 using the MRS D$-'$ PDF. Also shown (CR) is a phenomenological low-energy fit
 (from ref.~\protect\cite{Gavai95}).
}
\end{figure}
\begin{figure}
\caption{The $J/\psi$ longitudinal momentum distributions compared with
 $\bar{p} N$ (top) and $pN$ (bottom) data using two parametrizations of
 the PDF, MRS D$-'$ (solid) and GRV (dashed)
 (from ref.~\protect\cite{Gavai95}).
}
\end{figure}


\begin{thebibliography}{99}
\bibitem{NDE88}{
 P.\ Nason, S.\ Dawson and R.K.\ Ellis,
 Nucl.\ Phys.\ {\bf B303} (1988) 607 and {\bf B327} (1988) 49;
\hfill\\
 W.\ Beenakker, R.\ Meng, G.A.\ Schuler, J.\ Smith and W.L.\ van Neerven,
 Nucl.\ Phys.\ {\bf B351} (1991) 507 and
 Phys.\ Rev.\ {\bf D40} (1989) 54}
\bibitem{EN89}
 {R.K.\ Ellis and P.\ Nason,
 Nucl.\ Phys.\ {\bf B312} (1989) 551;
\hfill\\
J.\ Smith and W.L.\ van Neerven,
 Nucl.\ Phys.\ {\bf B374} (1992) 36}
\bibitem{LRSN93}
 {E.\ Laenen, S.\ Riemersma, J.\ Smith and W.L.\ van Neerven,
 Nucl.\ Phys.\ {\bf B392} (1993) 162; ibid.\ 229}
\bibitem{DKZZ93}{
 M.\ Drees, M.\ Kr\"amer, J.\ Zunft and P.M.\ Zerwas,
 Phys.\ Lett.\ {\bf B306} (1993) 371}
\bibitem{LRSN94}
 {E.\ Laenen, S.\ Riemersma, J.\ Smith and W.L.\ van Neerven,
 Phys.\ Rev. {\bf D49} (1994) 5753}
\bibitem{FMNR93}{
 S.\ Frixione, M.L.\ Mangano, P.\ Nason and G.\ Ridolfi,
 Nucl.\ Phys.\ {\bf B412} (1994) 225}
\bibitem{Harris}{
B.W.\ Harris and J.\ Smith, Stony Brook preprint ITP-SB-94-06,
January 1995 (hep-ph/9502312) and ITP-SB-95-08 in preparation}
\bibitem{FMNR94}{
 S.\ Frixione, M.L.\ Mangano, P.\ Nason and G.\ Ridolfi,
 Nucl.\ Phys.\ {\bf B431} (1994) 453}
\bibitem{HMRS}{
 P.\ Harriman, A.D.\ Martin, R.G.\ Roberts and W.J.\ Stirling,
 \PR {\bf D37} (1990) 798;
\hfill\\
 P.J.\ Sutton, A.D.\ Martin, R.G.\ Roberts and W.J.\ Stirling,
 \PR {\bf D45} (1992) 2349}
\bibitem{Mcgaughey}{
 R.V.\ Gavai, S.\ Gupta, P.L.\ McGaughey, E.\ Quack,
  P.V.\ Ruuskanen, R.\ Vogt and X.-N.\ Wang,
  Darmstadt preprint GSI-94-76, November 1994}
\bibitem{GRV}{
 M. Gl\"{u}ck, E. Reya and A. Vogt, Z. Phys. {\bf C53} (1992) 127}
\bibitem{D0}{
 A.D.~Martin, W.J.~Stirling and R.G.~Roberts,
 Phys. Lett. {\bf B306} (1993) 145}
\bibitem{KEK}{
 TOPAZ collaboration,
 R.\ Enomoto et al., Phys.\ Lett.\ {\bf B328} (1994) 535,
 Phys.\ Rev. {\bf D50} (1994) 1879;
\hfill\\
 VENUS collaboration, S.\ Uehara et al., Z.\ Phys.\ {\bf C63} (1994) 213}
\bibitem{Aleph}{
 A.\ Finch for the ALEPH collaboration, in Proc.\ of the Workshop
 on Two-Photon Physics at LEP and HERA, eds.\ G.\ Jarlskog and L.\
 J\"onsson, Lund, Sweden, May 1994}
\bibitem{Brodsky92}{
 S.J.\ Brodsky, P.\ Hoyer, A.H.\ Mueller and W.-K.\ Tang,
 Nucl.\ Phys.\ {\bf B369} (1992) 519}
\bibitem{HERA}{
 I.\ Abt et al.\ (H1), \NP {\bf B407} (1993) 515;
\hfill\\
 M.\ Derrick et al.\ (Zeus), \PL {\bf B316} (1993) 412}
\bibitem{MRSA}A.D.~Martin, W.J.~Stirling and R.G.~Roberts,
  Phys.\ Rev.\ {\bf D50} (1994) 6734
\bibitem{Einhorn}{M.B.\ Einhorn and S.D.\ Ellis,
 \PR {\bf D12} (1975) 2007;
\hfill\\
H.\ Fritzsch, \PL {\bf B67} (1977) 217;
\hfill\\
M.\ Gl\"uck, J.F.\ Owens and E.\ Reya, \PR {\bf D17} (1978) 2324;
\hfill\\
J.\ Babcock, D.\ Sivers and S.\ Wolfram, \PR {\bf D18} (1978) 162}
\bibitem{Schuler}
{For a recent review, see G.A.\ Schuler, ``Quarkonium production and
decays", preprint CERN-TH.7170/94, February 1994,
 to appear in Phys.\ Rep.}
\bibitem{CS}{
C.H.\ Chang, \NP {\bf B172} (1980) 425;
\hfill\\
E.L.\ Berger and D.\ Jones, \PR {\bf D23} (1981) 1521;
\hfill\\
R.\ Baier and R.\ R\"uckl, \PL {\bf B102} (1981) 364
  and \ZP {\bf C19} (1983) 251;
\hfill\\
J.G.\ K\"orner, J.\ Cleymans, M.\ Kuroda and G.J.\ Gounaris,
 \PL {\bf B114} (1982) 195 and Nucl.\ Phys.\ {\bf B204} (1982) 6}
\bibitem{Kuhn81}{
 J.H.\ K\"uhn and H.\ Schneider, \PR {\bf D24} (1981) 2996 and
 \ZP {\bf C11} (1981) 263}
\bibitem{Braaten93}{
 E.\ Braaten and T.C.\ Yuan, Phys.\ Rev.\ Lett.\ {\bf 71} (1993) 1673}
\bibitem{BCY93}{
 E.\ Braaten, K.\ Cheung and T.C.\ Yuan , \PR {\bf D48} (1993) 4230}
\bibitem{Braaten94}{
 E.\ Braaten and T.C.\ Yuan, \PR {\bf D50} (1994) 3176}
\bibitem{Chen93}{
 Y.Q.\ Chen, \PR {\bf D48} (1993) 5181;
\hfill\\
T.C.\ Yuan, \PR {\bf D50} (1994) 5664}
\bibitem{BDFM94}{
 E.\ Braaten, M.A.\ Doncheski, S.\ Fleming and M.L.\ Mangano,
 Phys.\ Lett.\ {\bf B333} (1994) 548}
\bibitem{Greco94}{
 M.\ Cacciari and M.\ Greco, Phys.\ Rev.\ Lett.\ {\bf 73}
  (1994) 1586;
\hfill\\
 D.P.\ Roy and K.\ Sridhar, Phys.\ Lett.\ {\bf B339} (1994) 141}
\bibitem{Sridhar94}{
  D.P.\ Roy and K.\ Sridhar,
  Phys.\ Lett.\ {\bf B345} (1995) 537}
\bibitem{Brodsky94}{
 M.\ V\"anttinen, P.\ Hoyer, S.J.\ Brodsky and W.-K.\ Tang,
 preprint SLAC-PUB-6637, August 1994 (hep-ph-9410237)}
\bibitem{Gavai95}{
 R.\ Gavai, D.\ Kharzeev, H.\ Satz, G.A.\ Schuler, K.\ Sridhar and R.\ Vogt,
 preprint CERN-TH.7526/94, December 1994 (hep-ph-9502270)}
\bibitem{Bodwin92}{
 G.T.\ Bodwin, E.\ Braaten, T.C.\ Yuan and G.P.\ Lepage,
 \PR {\bf D46} (1992) R3703}
\bibitem{Mannel}
  T.\ Mannel and G.A.\ Schuler, preprints
  CERN-TH.7468/94, September 1994 (hep-ph/9410333), Z.\ Phys.\ C in press,
 and CERN-TH.7523/94, December 1994 (hep-ph/9412337),
 Phys.\ Lett.\ B in press
\bibitem{NRQCD}
  G.T.\ Bodwin, E.\ Braaten and G.P.\ Lepage,
  Phys.\ Rev.\ {\bf D51} (1995) 1125
\bibitem{Fleming94}{
  E.\ Braaten and S.\ Fleming, Northwestern Univ.\ preprint
  NUHEP-TH-94-26, November 1994 (hep-ph/9411365)}
\bibitem{private}{
  V.\ Papadimitriou for the CDF collaboration,
  ``Production of heavy quark states in CDF", talk presented at the
  XXXth. Rencontre de Moriond, Les Arcs, Savoie, France, March 1995}
\bibitem{Cho}{
 P.\ Cho and M.B.\ Wise, Phys.\ Lett.\ {\bf B346} (1995) 129}

\end{thebibliography}
\end{document}